\documentclass{aastex6}
\usepackage{amssymb,amsmath}

\def\kms{km~s$^{-1}$}
\def\teff{$T_\mathrm{eff}$}

\newcommand{\vsini}{\ensuremath{v_{{\mathrm e}}\sin i}}

\fullcollaborationName{The Friends of AASTeX Collaboration}

\begin{document}


\title{HR 8844: a new transition object between the Am stars and the HgMn stars ?}

\author{R. Monier\altaffilmark{1}}
\affil{LESIA, UMR 8109, Observatoire de Paris et Universit\'e Pierre et Marie Curie Sorbonne Universit\'es, place J. Janssen, Meudon.}

\author{M. Gebran\altaffilmark{2}}
\affil{Department of Physics and Astronomy, Notre Dame University-Louaize, PO Box 72, Zouk Mikael, Lebanon.}

\author{F. Royer\altaffilmark{3}}
\affil{GEPI, Observatoire de Paris, place J. Janssen, Meudon, France.}

\and
\author{T. Kilicoglu\altaffilmark{4}}
\affil{Department of Astronomy and Space Sciences, Faculty of Science, Ankara University, 06100, Turkey.}

\and
\author{Y. Fr\'emat\altaffilmark{5}}
\affil{Royal observatory of Belgium, Dept. Astronomy and Astrophysics, Brussels, 8510, Belgium.}



\begin{abstract}

While monitoring a sample of apparently slowly rotating superficially normal early A stars, we have discovered that HR 8844 (A0 V), is actually  a new Chemically Peculiar star.
We have first compared the high resolution spectrum of HR 8844 to that of four slow rotators near A0V ($\nu$ Cap, $\nu$ Cnc , Sirius A and HD 72660) to highlight similarities and differences.
The lines of  \ion{Ti}{2}, \ion{Cr}{2},   \ion{Sr}{2} and \ion{Ba}{2} are conspicuous features  in the high resolution high signal-to-noise SOPHIE spectra of HR 8844 and much stronger than in the spectra of the normal star $\nu$ Cap. The \ion{Hg}{2} line at 3983.93 \AA\ is also  present in a 3.5 \% blend. 
Selected unblended lines of 31 chemical elements from He up to Hg  have been synthesized using model atmospheres computed with ATLAS9 and the spectrum synthesis code SYNSPEC48 including  hyperfine structure of various isotopes when relevant. These synthetic spectra have been  adjusted to the  mean SOPHIE spectrum of HR 8844,  and high resolution spectra of the comparison stars.
Chisquares were minimized in order to derive abundances or upper limits to the abundances of these  elements for HR 8844 and the comparison stars. 
HR 8844 is found to have underabundances of He, C, O, Mg, Ca and Sc, mild enhancements of Ti, V, Cr, Mn and distinct enhancements of the heavy elements  Sr, Y, Zr, Ba, La, Pr, Sm, Eu and Hg, the overabundances increasing steadily with atomic number.  
 This chemical pattern suggests that HR 8844 may actually be a new transition
  object between the coolest HgMn stars and the Am stars.
\end{abstract}





\section{Introduction} \label{sec:intro}

 A bibliographic query of the CDS for HR 8844 actually reveals only 34 publications.  HR 8844 appears in \citeauthor{Cowley1969}'s classification of the bright A stars and was ascribed an A0V spectral type by these authors. They did not comment on any peculiarity of the spectrum. HR 8844 is also in Eggen's survey of A0 stars \citep{1984ApJS...55..597E}.
 It appears as a single star in \cite{2008MNRAS.389..869E}.
 The purpose of this paper is to perform a detailed abundance analysis of HR 8844 and show that this star is a new Chemically Peculiar (CP)  star. \\
 We have recently undertaken a spectroscopic survey of all apparently slowly rotating bright early A stars (A0-A1V) and late B stars (B8-B9V) observable from the northern hemisphere. 
This project addresses fundamental questions of the physics of late-B and early-A stars: i) can we find new instances of rapid rotators seen pole-on (other than Vega) and study their physical properties (gradient of temperature across the disk, limb and gravity darkening), ii) is our census of Chemically Peculiar stars complete up to the magnitude limits we adopted ? If not, what are the  physical properties of the newly found CP stars? 
   The abundance results for the A0-A1V sample have been published in \cite{Royer}.
Targets have been observed with SOPHIE, the \'echelle high-resolution spectrograph at Observatoire de Haute Provence yielding spectra covering the 3900--6800\,\AA\ spectral range over 39 orders at a resolving power $R = 75000$.
   A careful abundance analysis of the high resolution high signal-to-noise ratio spectra of the A stars sample and a hierarchical classification have allowed to sort out the sample of 47 A stars into 17 chemically normal stars (ie. whose abundances do not depart by more  than $\pm 0.20$ dex from solar values), 12 spectroscopic binaries and 13 Chemically Peculiar stars (CPs) among which five are new CP stars. The status of these new CP stars still needs to be fully specified by spectropolarimetric observations to address their magnetic nature or by exploring new spectral ranges which we had not explored in this first study. 
   \\
   We have now started to examine the B8-B9V sample using the full wavelength coverage provided by SOPHIE to search for new Chemically Peculiar stars. We have already reported on the discovery of 4 new HgMn stars \citep{Monier} whose spectra display strong 
 \ion{Hg}{2} lines at 3984\,\AA\ and strong \ion{Mn}{2} lines. In the process of our analysis of the B8-B9V sample, we also found that HD 67044 is most likely another new HgMn star \citep{Monier2016}.
\\
\cite{Royer} performed abundance analyses for C, O, Mg, Si, Ca, Sc, Ti, Cr, Fe, Sr, Y, Zr using Takeda's automated procedure and classified HR 8844 as ``probably normal'': the four criteria used for the automatic classification were not fully consistent, depending on the chemical species used.
 In order to clarify the nature of HR 8844, we have compared its spectra with high resolution high signal-to-noise spectra of 4 stars near A0V and B9V: the superficially normal $\nu$ Cap (B9V) and the cool HgMn star $\nu$ Cnc, Sirius A (A1m) and HD 72660 (A1m).
 We  have then determined the abundances of 31 chemical elements, in particular helium and several species heavier than barium (not studied in \citealt{Royer}), for HR 8844 using spectrum synthesis to quantify the enhancements and depletions of these elements or to provide upper limits.  
 We also performed the abundance analysis for the four comparison stars using as much as feasible the same lines and atomic data as employed for HR 8844 for consistency in order to compare the chemical composition of HR 8844 with that of these comparison stars and clarify the nature of this object. 

 \section{Observations and reduction}
\label{sec:obs-reduc}

We obtained one spectrum of HR 8844 on 05 August 2009 and then secured 3 new spectra in December 2016 at Observatoire de Haute Provence using the high resolution ($R = 75000$) mode of the SOPHIE \'echelle spectrograph \citep{Perruchot}.
  The  $\frac{S}{N}$ ratio of the individual spectra ranges from  251 up to  381 at 5500 \AA\ . 
  For $\nu$ Cap, we secured four high resolution spectra with HERMES \citep{2011A&A...526A..69R} at the Roque de los Muchachos Observatory. 
For the three other comparison stars, we fetched spectra from spectroscopic archives. For HD 72660, we fetched a spectrum from the HARPS archive (R=125000) and for Sirius A several I profiles obtained with NARVAL from the the Polarbase database (R=75000). For $\nu$ Cnc, we have retrieved a spectrum form the ELODIE archive (R=45000).
 The observations log  of these data for the 5 stars is displayed in Table\,\ref{table:1}.
 
\begin{table}
\tiny
\caption{Observation log for HR 8844, Vega, $\nu$ Cap, $\nu$ Cnc, Sirius A and HD72660} 
\label{table:1} 
\centering 
\begin{tabular}{cccccccccc} 
\hline\hline 
Star ID & Spectral & V &Observation & Instrument & Resolving & Exposure & S/N & S/N & S/N \\ 
          & type        &  & Date             &                   & power         &    time (s ) & at 3900 \AA\ & at 5000 \AA\ & at 6000 \AA\     \\ 
\hline 
HR 8844 & A0V    & 5.89 & 2009-08-05 & SOPHIE & 75000 & 600 & 143 & 269 & 274\\ 
 HR 8844 & A0V    & 5.89 & 2016-12-12& SOPHIE & 75000 & 900 & 202 & 381 & 389 \\ 
 HR 8844 & A0V    & 5.89 & 2016-12-13 & SOPHIE & 75000 & 400  & 133 & 251 & 256 \\     
 HR 8844 & A0V    & 5.89 & 2016-12-14 & SOPHIE & 75000 & 600  & 174 & 328 & 335 \\ 
Vega   & A0V       & 0.00  & 2012-08-06 & SOPHIE & 75000 & 25   & 309 & 583 & 595   \\          
$\nu$ Cap & B9IV & 4.76 & 2014-08-16  & HERMES & 85000  & 150       & 85 & 160 & 163   \\
$\nu$ Cap & B9IV & 4.76 & 2014-08-16  & HERMES & 85000  & 150       & 54 & 102 & 104  \\
$\nu$ Cap & B9IV & 4.76 & 2014-08-16  & HERMES & 85000  & 150       & 87 & 164 & 167  \\
$\nu$ Cap & B9IV & 4.76 & 2014-08-16  & HERMES & 85000  & 150       & 80 & 150 &153  \\
$\nu$ Cnc  &B9.5VHgMn &  5.45 & 2005-02-05 & ELODIE & 42000  & 3600 & 182 & 344 & 350 \\
Sirius A &A1Vm  &-1.46 & 2007-03-12  & NARVAL  & 75000 & 2    & 239 & 450  & 459    \\
HD 72660 & A0Vm &  5.80 & 2012-02-19 &HARPS    & 125000 & 117.8     & 77  & 146 & 149 \\
\hline 
\end{tabular}
\end{table}

   The SOPHIE, ELODIE, HARPS, NARVAL and HERMES data are automatically reduced by the individual projects  to produce 1D extracted and wavelength calibrated \'echelle orders. Each reduced order was normalised separately using a Chebychev polynomial fit with sigma clipping, rejecting points above or below 1 $\sigma$ of the local continuum. Normalized orders were then  merged together, corrected  by the blaze function and resampled into a constant wavelength step of about 0.02\,\AA\ \cite[see][for more details]{Royer}.
 The radial velocity of HR 8844, $\nu$ Cap and HD 72660  were derived in \cite{Royer} using cross-correlation techniques, avoiding the Balmer lines and the atmospheric telluric lines. The normalized spectrum was cross-correlated with a synthetic template extracted from the POLLUX database\footnote{\url{http://pollux.graal.univ-montp2.fr}}  \citep{Palacios} corresponding to the parameters $T_\mathrm{eff}=11000$\,K, $\log g=4$ and solar abundances. A parabolic fit of the upper part of the resulting cross-correlation function yields the Doppler shift, which is then used to shift spectra to rest wavelengths. The projected rotational velocities are taken from \cite{Royer} who derived them from the position of the first zero of the Fourier transform of individual lines.   The radial velocity and projected equatorial velocity of HR 8844, $\nu$ Cap, HD 72660, $\nu$ Cap and $\nu$ Cnc are collected in Table 2. For Sirius A, they were retrieved from \cite{land2011}.

\section{The line spectrum of HR 8844 - Comparison to $\nu$ Cap}\label{sec:linespec}

\subsection{The line spectrum of HR 8844}

In order  to establish the chemical peculiarity of HR 8844, we have measured the centroids of all 552 lines absorbing more than 2\% of the continuum. Almost all these lines could be identified after we performed a complete synthesis of the spectrum and derived the abundances. The proposed identifications corresponding to the final abundances of HR 8844 are collected in Table 3.
The strongest lines (ie. which absorb more than 5\% of the continuum)  are due to \ion{Mg}{2}, \ion{Mg}{1}, \ion{Ca}{2} (resonance lines), \ion{Ca}{1} (idem), \ion{Ti}{2},
\ion{Cr}{2}, \ion{Fe}{2}, \ion{Fe}{1}, \ion{Sr}{2} and \ion{Ba}{2}.
 Several spectral regions have been inspected to search for the chemical peculiarity of HR 8844. First, the red wing of H${\epsilon}$, harbors a line close to the location of the \ion{Hg}{2} $\lambda$ 3983.93\,\AA\ line and several \ion{Zr}{2} and \ion{Y}{2} lines likely to be strengthened in CP stars. After proper correction for the stellar radial velocity, we found that HR 8844
does  show a feature next to the \ion{Hg}{2} 3983.93\,\AA\ line absorbing about 3.5 \% , and also the lines of \ion{Y}{2} at 3982.59 \AA\ (2 \%) and of \ion{Zr}{2} at 3991.13 \AA\ (3.5 \%) and 3998.97 \AA\ (2 \%).
Several other lines of \ion{Y}{2} and \ion{Zr}{2} could be identified in the spectrum of HR 8844 and have been used to derive the abundances of these elements.
Second, we examined the region from 4125\,\AA\ to 4145\,\AA\  for the \ion{Si}{2} lines at 4128.054 \AA\ and 4130.894 \AA\ 
and  the  \ion{Mn}{2} line at 4136.92\,\AA. 
This \ion{Mn}{2} line is definitely present in HR 8844 but is only a weak line (0.5 \%). Similarly, the  lines of \ion{Mn}{2} at 4206.37\,\AA\ and 4252.96\,\AA\ absorb respectively 1.5 \% and 2.0 \%.
The abundance analysis of two \ion{Mn}{2} lines with hyperfine structure will reveal a manganese excess of about 2 times the solar value.

\subsection{Comparison to spectra of $\nu$ Cap}

\begin{figure}[h!]
\vskip 0.5cm
   \centering
      \includegraphics[scale=0.35]{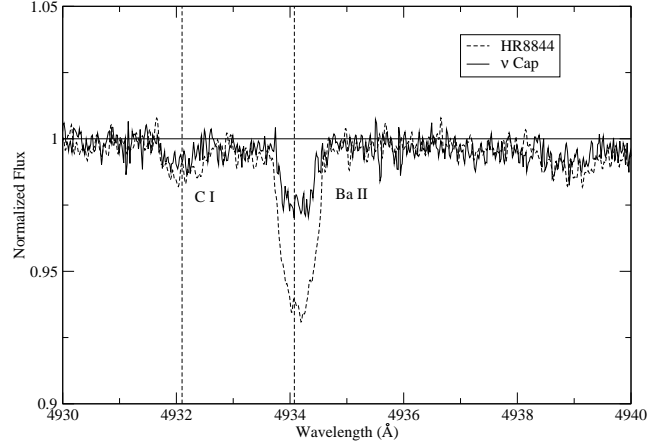}
   \caption{Comparison of the \ion{C}{1} lines and the \ion{Ba}{2} resonance line at 4934.076 \AA\ in the spectra of $\nu$ Cap (solid line) and HR 8844 (dashed lines) }
   \label{figure1}
\end{figure}

\begin{figure}[h!]
\vskip 0.5cm
   \centering
      \includegraphics[scale=0.35]{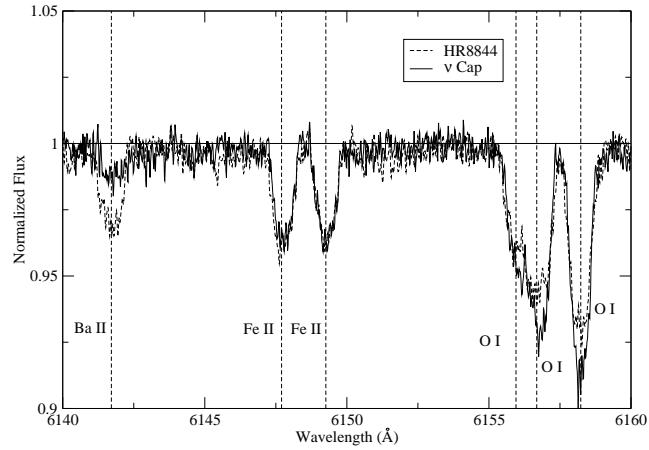}
   \caption{Comparison of the \ion{O}{1} lines at 6155.87, 6156.62 and 6158.17 \AA, the \ion{Ba}{2} line at 6141.713 \AA\ in the spectra of $\nu$ Cap (solid line) and HR 8844 (dashed lines) }
   \label{figure2}
\end{figure}

\begin{figure}[h!]
\vskip 0.5cm
   \centering
      \includegraphics[scale=0.35]{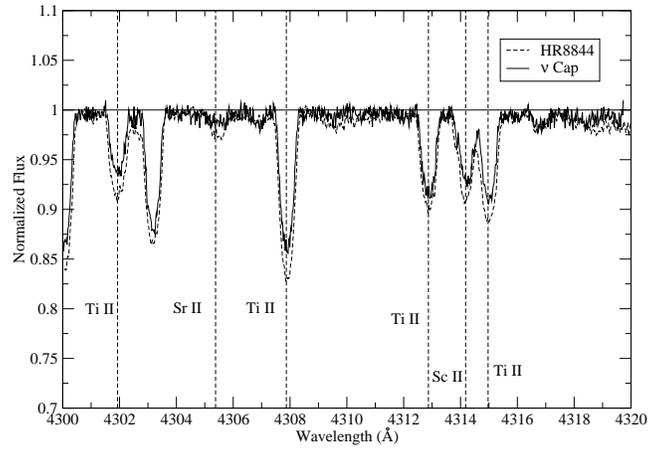}
   \caption{Comparison of four \ion{Ti}{2} lines, the \ion{Sr}{2} line at 4305.443 \AA\ and the  \ion{Sc}{2} line at 4314.18 \AA\ in the spectra of $\nu$ Cap (solid line) and HR 8844 (dashed lines) }
   \label{figure3}
\end{figure}

\begin{figure}[h!]
\vskip 0.5cm
   \centering
      \includegraphics[scale=0.35]{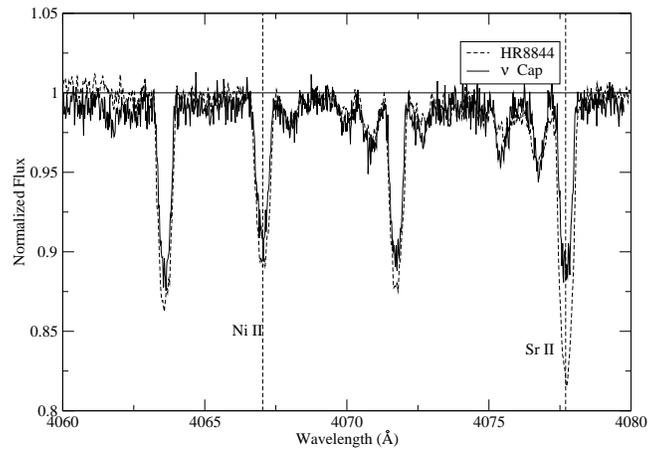}
   \caption{Comparison of the \ion{Ni}{2} lines at 4067.04 \AA, the \ion{Sr}{2} resonance line at 4077.70 \AA\ in the spectra of $\nu$ Cap (solid line) and HR 8844 (dashed lines) }
   \label{figure4}
\end{figure}

\begin{figure}[h!]
\vskip 0.5cm
   \centering
      \includegraphics[scale=0.35]{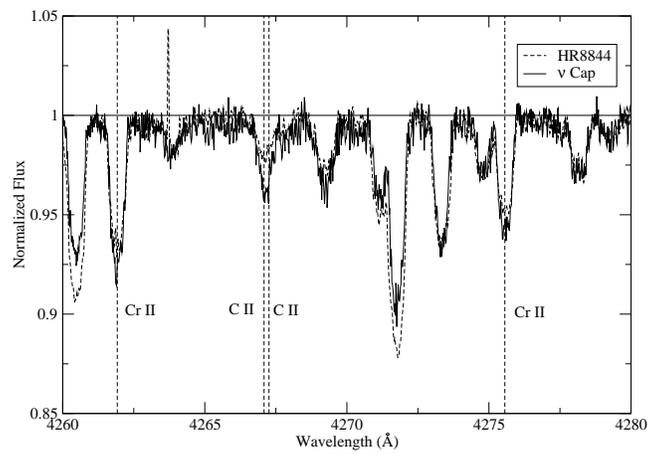}
   \caption{Comparison of the  \ion{C}{2} doublet at 4267.00 and 4267.26 \AA\ and the \ion{Cr}{2} lines at 4261.92 and 4275.56 \AA\ in the spectra of $\nu$ Cap (solid line) and HR 8844 (dashed lines) }
   \label{figure5}
\end{figure}

HR 8844, Vega and $\nu$ Cap have similar effective temperatures, surface gravities and projected equatorial velocities \vsini.
Differences in the line intensities should therefore reflect mostly differences in chemical compositions.
Several spectral regions highlight the differences between HR 8844 and $\nu$ Cap.
Several lines are stronger in HR 8844 than in $\nu$ Cap and Vega which shows than HR 8844 is enriched in the elements responsible for these lines with respect to $\nu$ Cap and Vega . We will only show the comparisons with $\nu$ Cap here.
Figure \,\ref{figure1} displays the comparison of the spectra of $\nu$ Cap
and HR 8844 in the range from 4930 \AA\ to 4940 \AA\ where three lines of \ion{C}{1} and the resonance line of \ion{Ba}{2}
at 4934.096 \AA\ fall. All lines of \ion{C}{1} are slightly stronger and the \ion{Ba}{2} line is much stronger in HR 8844.
Figure \,\ref{figure2} displays the comparison of the spectra of $\nu$ Cap
and HR 8844 in the range 6140 to 6160 \AA\  where several lines of \ion{O}{1} fall together with the \ion{Fe}{2} lines at 6147.62 \AA\ and
 6149.26 \AA\ and the \ion{Ba}{2} line at 
6141.713 \AA. All \ion{O}{1} lines are weaker in HR 8844 whereas the \ion{Ba}{2} line is much stronger in HR 8844.
Figure \,\ref{figure3} displays the comparison of the spectra of $\nu$ Cap
and HR 8844 in the range 4300 to 4320 \AA\ where 4 \ion{Ti}{2} lines, the \ion{Sc}{2} line at 4314.18 \AA\ and the \ion{Sr}{2}
line at 4305.443 \AA\ fall. The lines of \ion{Ti}{2}, \ion{Sc}{2} and \ion{Sr}{2} are all stronger in the spectrum of HR 8844 than in that of $\nu$ Cap.
Figure \,\ref{figure4} displays the comparison of the spectra of $\nu$ Cap
and HR 8844 in the range 4060 \AA\ up to 4080 \AA\ where the \ion{Ni}{2} at 4067.04 \AA\ and the \ion{Sr}{2} resonance line at 4077.70 \AA\ fall.
These two lines are stronger in the spectrum of HR 8844 than in that of $\nu$ Cap.
Finally, Fig. \,\ref{figure5} displays the comparison of the SOPHIE spectra of $\nu$ Cap
and HR 8844 in the range 4260 up to 4280 \AA\ where the  \ion{C}{2} doublet at 4267.00 \AA\ and 4267.26 \AA\ and the \ion{Cr}{2} lines at 4261.92 \AA\ and 4275.56 \AA\ fall. The \ion{C}{2} lines are weaker in HR 8844 than in $\nu$ Cap and the \ion{Cr}{2} lines are slightly stronger in HR 8844. 
From these first comparisons, we infer that C, O should be less abundant in HR 8844 than in $\nu$ Cap, whereas Fe may be comparable and Ti, Cr, Ni, Sr and Ba are more abundant in HR 8844 than in $\nu$ Cap. Similar trends are found when comparing the spectra of Vega and HR 8844.
The abundance analysis carried out in next paragraph will confirm this.

\section{Abundance determinations for HR 8844, $\nu$ Cap, $\nu$ Cnc, Sirius A and HD 72660}
 
 \subsection{Fundamental parameters determinations}
 
For HR 8844 , $\nu$ Cap and $\nu$ Cnc, we have applied Napiwotzky's (\citeyear{Napiwotzki}) UVBYBETA procedure to derive the effective temperature and surface gravity. The fundamental parameters of Sirius A and HD~72660 were taken respectively from \cite{land2011} and \cite{GL2016}. 
The adopted effective temperatures, surface gravities, projected equatorial velocities and radial velocities for HR 8844, $Vega, \nu$ Cap, $\nu$ Cnc, Sirius A and HD 72660
 are collected in Table\,\ref{table:2}. The adopted values for the parameters of HR8844 are in good agreement with the spectroscopic determination of \cite{PCA}. The projected equatorial velocities and radial velocities of HR 8844 and HD 72660 are taken from \cite{Royer} and for $\nu$ Cap from \cite{2002A&A...393..897R}.

 \begin{table}
\caption{Adopted fundamental parameters for HR 8844, Vega, $\nu$ Cap, $\nu$ Cnc, Sirius A, HD 72660}   
\label{table:2}      
\centering                          
\begin{tabular}{c c c c cc}        
\hline\hline                 
Star ID & $T_\mathrm{eff}$ & $\log g$  & $v\sin i$ & $V_{rad}$ & $\xi$\\    
            &   (K)            &               &  (km\,s$^{-1}$) & (km\,s$^{-1}$) & (km\,s$^{-1}$) \\
\hline                        
   HR 8844 & 9752 $\pm$ 250 & 3.80 $\pm$ 0.25  & 27.3 $\pm$ 0.3 & -4.48 $\pm$ 0.21& 1.40  $\pm$ 0.20 \\  
   Vega       & 9500 $\pm$ 250  & 4.00 $\pm$ 0.25 & 27.0  $\pm$ 0.3  & -17.51 $\pm$ 0.02 & 1.70 $\pm$ 0.30 \\
   $\nu$ Cap &10300 $\pm$ 250 &3.90 $\pm$ 0.25 & 24.0 $\pm$ 0.3  & -11.39 $\pm$ 0.20 & 0.50 $\pm$ 0.20  \\
   $\nu$ Cnc & 10300 $\pm$ 250 &3.67 $\pm$ 0.25 & 18.0 $\pm$ 0.3   & -20.04 $\pm$ 0.20  &0.10 $\pm$ 0.20   \\
   Sirius A   &9900 $\pm$ 250  &4.30 $\pm$ 0.25 & 16.5 $\pm$ 0.3 &  variable           & 2.10 $\pm$ 0.30       \\
   HD 72660 &9650 $\pm$ 250 &4.05 $\pm$ 0.25 & 5.0 $\pm$ 0.5      & 4.50 $\pm$ 0.07   & 2.20 $\pm$ 0.20  \\
\hline    \\                               
\end{tabular}
\end{table}

 \subsection{Microturbulent velocity determination}
 
\begin{figure}[h!]
\vskip 0.5cm
   \centering
      \includegraphics[scale=0.35]{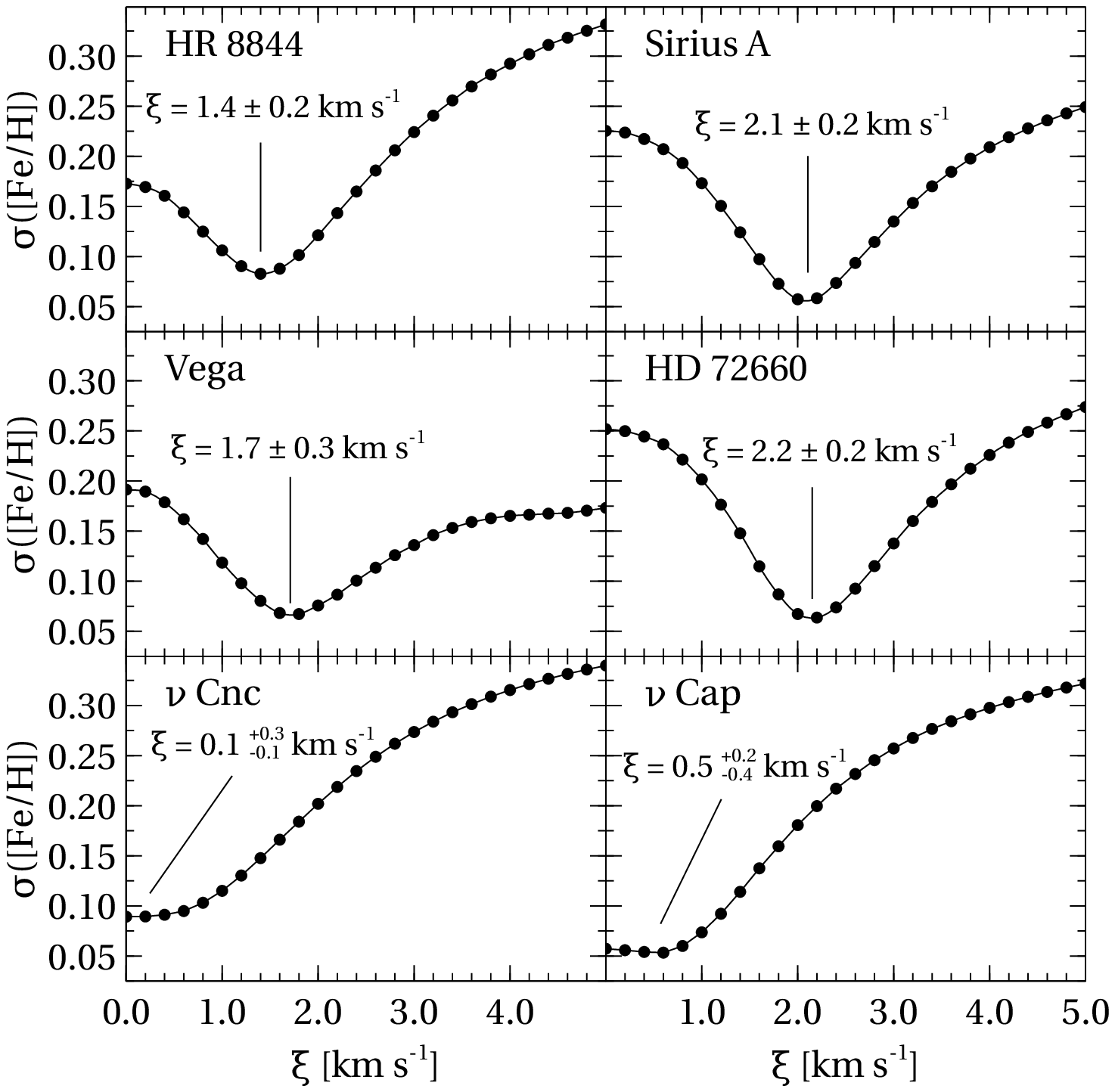}
      \caption{Determination of the microturbulent velocities of HR 8844, Vega, $\nu$ Cap, $\nu$ Cnc, Sirius A and HD 72660.}
        \label{vmicr}
 \end{figure}

In order to derive the microturbulenct velocity of HR 8844, Vega, $\nu$ Cap, $\nu$ Cnc, Sirius A and HD 72660, we have simultaneously derived the iron abundance
[Fe/H] for 50 unblended \ion{Fe}{2} lines and  a set of microturbulent
velocities ranging from 0.0 to 5.0 \kms. Figure \,\ref{vmicr} shows the standard
deviation of the derived [Fe/H] as a function of the microturbulent velocity.
The adopted microturbulent velocities are the values which minimize the standard deviations ie. for that value, all \ion{Fe}{2} lines yield
the same iron abundance.
We therefore adopt a microturbulent velocity $\xi_t$ = 1.4 $\pm$ 0.2 \kms\ constant with depth  for HR 8844. The microturbulent velocities of the six stars are collected in Table 2.

 \subsection{Model atmospheres and spectrum synthesis calculations}
 
 Plane parallel model atmospheres assuming radiative equilibrium and hydrostatic equilibrium were computed using the ATLAS9 code \citep{Kurucz} for the appropriate fundamental parameters of each star. The linelist was built from Kurucz's gfall21oct16.dat  \footnote{\url{http://kurucz.harvard.edu/linelists/gfnew/gfall21oct16.dat}} which includes hyperfine splitting levels.
The observed wavelengths and oscillator strengths were retrieved by querying the NIST\footnote{\url{http://www.nist.gov/}
Atomic Spectra Database \citep{Kramida2017} for the following atoms or ions: \ion{He}{1}.
\ion{C}{2}, \ion{O}{1}, \ion{Mg}{2}, \ion{Al}{1}, \ion{Al}{2}, \ion{Si}{2}, \ion{S}{2}, \ion{Ca}{2}, \ion{Sc}{2}, \ion{Ti}{2}, \ion{Cr}{2}, \ion{Mn}{2},
\ion{Fe}{2}, \ion{Sr}{2}, \ion{Ba}{2} and \ion{Dy}{2}}.
For the other species which lack data in the Atomic Spectra Database, we have used other references to retrieve their atomic data
which are collected in Tab.~5.  
 A grid of synthetic spectra was computed with SYNSPEC48 \citep{Hubeny} to model the
 lines of 31 elements  for HR8844 and $\nu$ Cap. Computations were iterated varying the unknown abundance until minimization of the chi-square between the observed and synthetic spectrum was achieved. Abundances are derived for 34 elements for $\nu$ Cnc, 37 elements for Sirius A and 37 elements for HD~ 72660 because these three stars have significantly lower projected rotational velocities.
 
\subsection{The derived abundances of HR 8844}

\begin{figure}[h!]
\vskip 0.5cm
   \centering
      \includegraphics[scale=0.35]{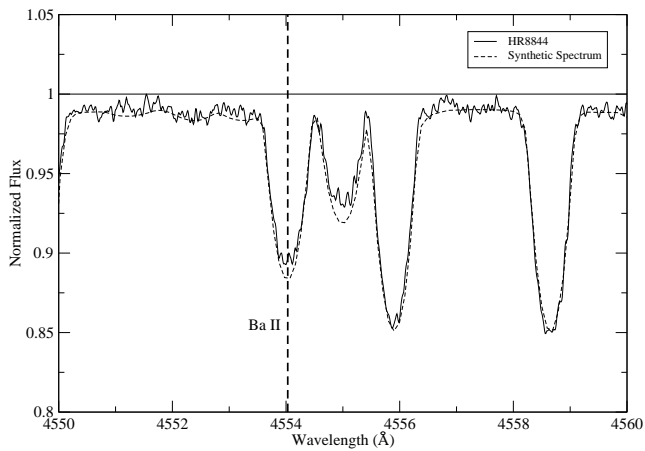}
      \caption{Determination of the barium overabundance from the \ion{Ba}{2} line at 4554.03 \AA\ in HR 8844}
        \label{comp_Ba}
 \end{figure}

We have used only unblended lines to derive the abundances.
For a given element, the final abundance is a weighted mean of the abundances derived for each transition (the weights are derived from the NIST grade assigned to that particular transition).
For several elements, in particular the heaviest elements, only one unblended line was available.
 These final abundances and their estimated uncertainties for HR 8844 
are collected in Tab.~4. The determination of the uncertainties is discussed in the Appendix.
Table 4 contains for each analysed species the adopted laboratory wavelength, logarithm of
oscillator strength, its source, the logarithm of the absolute abundance normalised to that of hydrogen (on a scale where $\log(N_{H}) = 12$) for each transition, and the solar abundance adopted for that element. 
In this work, we adopted \cite{1998SSRv...85..161G}  abundances for the Sun as a reference.
The final mean abundance and its estimated uncertainty are then given.
\\
The light elements, He, C , O, Mg are found to be  depleted in HR 8844. 
The helium  abundance is found to be about 0.70 times the solar abundance of helium consistently from the He I lines near 4471 \AA\ and near 5875.5 \AA.
The NLTE abundance correction for the $\lambda$ 4471.48 \AA\  \ion{He}{1} line is very small as shown by \cite{lemke} for early A-type stars and we have not corrected for it. 
Carbon is depleted, about 0.40 times the solar abundance from the analysis of the \ion{C}{2} triplet at 4267 \AA.
Oxygen is depleted about 0.80 times the solar oxygen abundance from 12 \ion{O}{1} transitions near 5330 \AA\ and from 6155 to 6158 \AA.
Magnesium is also depleted by 0.45 times the solar abundance, aluminium is 0.8 times solar and silicon is about solar. Calcium and scandium are depleted by 0.70 times the solar abundances.
As from titanium, all elements start to be mildly overabundant, except for iron. From the analysis of 12 lines, titanium is found to be about 1.41 times enriched respect to the solar abundance.
The vanadium abundance is 1.88 times the solar abundance and the chromium abundance 1.45 times the solar abundance.
The \ion{Mn}{2} lines at 4206.37 \AA\ and 4259.19 \AA\ include hyperfine structure of the isotopes of manganese and yield 2 times the solar manganese abundance.
 The iron abundance which is solar has been derived mostly by using 16 \ion{Fe}{2}  lines of multiplets 37, 38 and 186  in the range 4500--4600\,\AA\ whose atomic parameters are critically assessed in NIST (these are C+ and D quality lines). These lines are widely spaced and the continuum is fairly easy to trace in this spectral region. Their synthesis always yields consistent iron abundances from the various transitions with very little dispersion. The iron abundance is probably the most accurately determined of the abundances derived here. 
The Sr-Y-Zr triad is overabundant by modest amounts: about 5 times solar for strontium, about 5 times for yttrium and 5.8 times solar for zirconium. Barium is enhanced by a factor of 9.3 and the Lanthanides by factors of 15 up to 100 times solar. 
Figure \ref{comp_Ba} illustrates the detection of a 10 times solar overabundance for barium from the \ion{Ba}{2} line at 4554.03 \AA.\\

\begin{figure}[h!]
\vskip 0.5cm
   \centering
      \includegraphics[scale=0.5]{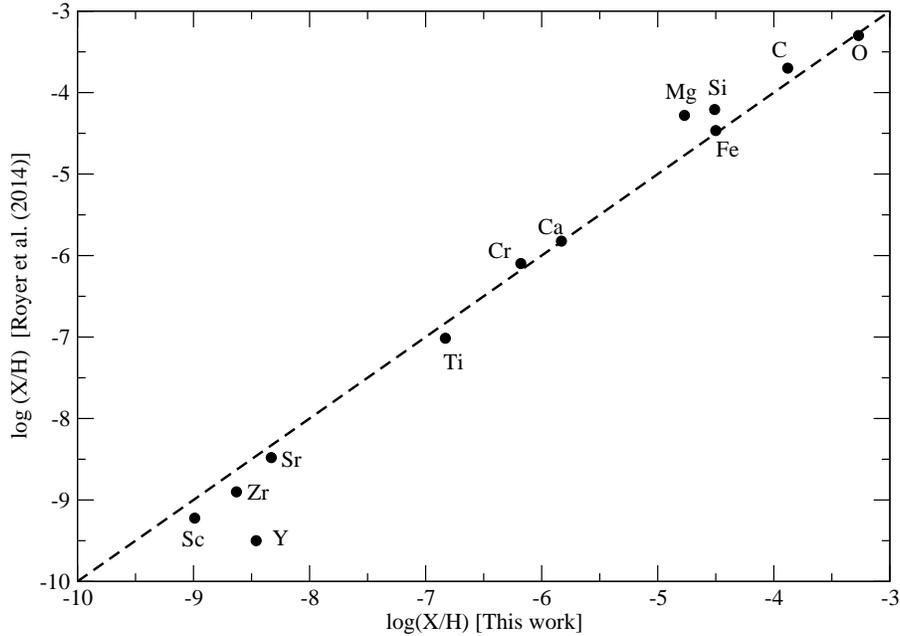}
      \caption{Comparison of the abundances derived in \cite{Royer} and this work for the twelve elements in common}
        \label{compabund}
 \end{figure}

Mercury is enhanced by 10000 times the solar value. 
 The overall abundance pattern of HR 8844 is therefore that light species tend to be almost all underabundant 
 (He, C, O, Mg, Si, Ca, Sc)  while the iron-peak elements show mild enhancements (less than 5 times solar) and Sr, Y, Zr, Ba, the Lanthanides  and Hg  show more and more pronounced overabundances (larger than 5 times solar), the largest overabundance being for Hg. The general trend therefore is that the heaviest elements are the most overabundant which strongly suggests that atomic diffusion be responsible for the chemical pattern of HR 8844.
 \\
 The derived abundances of HR 8844 are compared to the previous determination in \cite{Royer} for the twelve elements in common in Tab.~7 and figure \ref{compabund}. We find a reasonably good agreement (within $\pm$ 0.25 dex) for C, O, Ca, Sc, Cr, Fe and Sr.
 
\hskip 1cm

\subsection{The derived abundances for  the comparison stars $\nu$ Cap, $\nu$ CnC, Sirius A and HD 72660}

Using the same method and linelist, we have derived the elemental abundances of the normal star $\nu$ Cap, the cool HgMn star
$\nu$ Cnc and the hot Am stars Sirius A and HD 72660 in order to compare the found abundances of HR 8844 to those of these four stars.
The found abundances for these four comparison stars are collected in Tab.~5.\\
We have compared the derived abundances for these four stars with published abundance studies in the literature as displayed in Tab.~7.
We expect differences with previous analyses because of i) differences in the adopted $\vsini$ or $\xi$ and ii) differences in the lines used and their atomic data.
The found abundances for $\nu$ Cap are very close to solar for most elements in agreement with \cite{adelman1991} and \cite{smith1993}.
This star proves to be a very good normal comparison star. The abundances in \cite{adelman1991} differ slightly from ours for three reasons. Although he did use a similar effective temperature and surface gravity, Adelman adopted a null microturbulent velocity for $\nu$ Cap whereas we use 0.50 \kms. We did not necessarily use the same lines for a given element and the atomic data for the few lines have been upgraded to the latest values compiled in NIST.
\\
For $\nu$ Cnc, our abundances are almost all larger than those reported in \cite{adelman1989}. He did use fundamental parameters and a microturbulent velocity similar to ours, however he adopted a smaller $\vsini $= 13 \kms whereas we use 18 \kms. We therefore have to slightly enhance the abundances to reproduce the observed profiles. Note that we determine here for the first time the abundances of several Lanthanides in $\nu$ Cnc.
\\
For Sirius A, we have compared our results with the most recent determinations of \cite{land2011} for the elements up to Nickel and \cite{cow2016} for heavier elements.
We use the same effective temperature, surface gravity and a slightly lower microturbulent velocity (2.10 \kms\ rather than 2.20 \kms\ adopted by these authors).
In his paper, \cite{land2011} emphasizes that even for Sirius large differences in elemental abundances remain for C, Mg, Al, Si, S, Ca, V, Mn, Ni and even Fe from one author to another. We find good agreement (ie. within $\pm$ 0.20 dex) with Landstreet's (2011) abundances for O, Ca, Cr, Mn, Fe, Ni, Sr, Ba. For elements heavier than barium,
we find a fair agreement with \cite{cow2016} abundances for La, Pr, Nd, Sm and Hf within $\pm$ 0.20 dex of their determinations.
\\
For HD 72660, we have compared our abundances with those of \cite{GL2016} as we have used similar effective temperature, surface gravity, projected equatorial velocity and microturbulent velocity. We find a fair agreement with \cite{GL2016} for the abundances of several elements: C, Mg, Al, Si, S, Sc, Ti, V, Cr, Ni and Zr.

 \begin{figure}[h!]
\vskip 0.5cm
   \centering
      \includegraphics[scale=0.50]{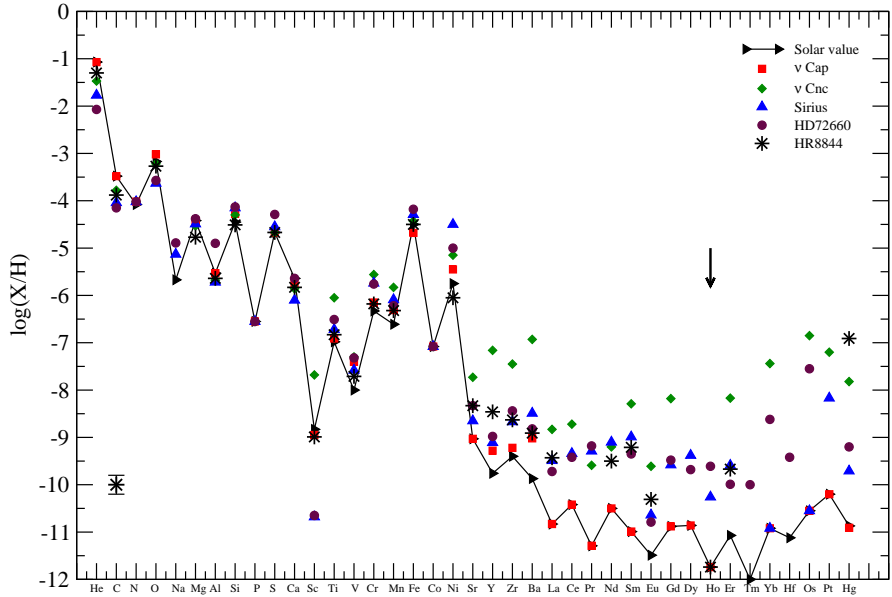}
      \caption{Comparison of the abundance pattern of HR 8844,  to those of  $\nu$ Cap, $\nu$ Cnc, Sirius A and HD 72660.}
        \label{comp_patterns}
 \end{figure}
 
 
In Fig.~\ref{comp_patterns}, we compare the found abundances for HR 8844 with our new determinations for the four comparsion stars.
As $\nu$ Cnc, and Sirius A , HR 8844 displays the characteristic underabundances for most elements lighter than titanium
and pronounced overabundances for elements heavier which Chemically Peculiar stars harbour. The overabundances of HR 8844 are modest for the iron-peak elements  but increase for the Sr-Y-Zr  triad and barium (of the order of 10 times solar) and for the Lanthanides  and of the order of $10^{2}$ solar for mercury. 
The overabundance of manganese and mercury, the underabundances of calcium and scandium and of the lightest elements , the mild overabundances of the iron-peak elements, of the Sr-Y-Zr triad, of barium and several Rare Earths show that HR 8844 is a new hot Am star which also displays characteristic enhancements of the coolest HgMn stars (as HD 72660 and Sirius A do). We therefore propose that HR 8844 be a new transition object between the Am stars and the coolest HgMn stars and as such is a very interesting star.

\section{Conclusion}

We report here on detailed abundance determinations of the A0V star HR 8844, hitherto considered as a "normal star". Our analysis definitely establishes that HR 8844
is a new Chemically Peculiar star, probably a hot Am star (as Sirius A and HD 72660 are) extending the realm of Am stars to significantly higher temperatures. As HD 72660 and Sirius A, 
HR 8844 may well be 
a transition object between the Am stars and the coolest HgMn star and as such is a very interesting object.
We are undertaking a monitoring of HR 8844 over its rotation period (which must be smaller than 5 days assuming a radius of about 2 solar radii for an A0V star) to search for variability due to the presence of spots or a companion.

\label{sec:conclusion}

\begin{appendix}
\section{Determination of uncertainties}
For a representative line of a given element, six major sources are included in the uncertainty determinations: the uncertainty on the effective temperature ($\sigma_{T_{\rm{eff}}}$), on the surface gravity ($\sigma_{\log g}$), on the microturbulent velocity ($\sigma_{\xi_{t}}$), on the apparent rotational velocity ($\sigma_{v_{e}\sin i}$), the oscillator strength ($\sigma_{\log gf}$) and the continuum placement ($\sigma_{cont}$). These uncertainties are supposed to be independent, so that the total uncertainty $\sigma_{tot_{i}}$ for a given transition (i) is:\\
\begin{equation}
\sigma_{tot_{i}}^{2}=\sigma_{T_{\rm{eff}}}^{2}+\sigma_{\log g}^{2}+\sigma_{\xi_{t}}^{2}+\sigma_{v_{e}\sin i}^{2}+\sigma_{\log gf}^{2}+\sigma_{cont}^{2}.
\end{equation} 
The mean abundance $<[\frac{X}{H}]>$ is then computed as a weighted mean of the individual abundances [X/H]$_{i}$ derived for each transition (i):\\
\begin{equation}
<[\frac{X}{H}]>=\frac{\sum_{i}([\frac{X}{H}]_{i}/\sigma_{tot_{i}}^{2})}{\sum_{i}(1/\sigma^{2}_{tot_{i}})}
\end{equation}
and the total error, $\sigma$ is given by: \\
\begin{equation}
\frac{1}{\sigma^{2}}=\sum_{i=1}^{N}(1/\sigma_{tot_{i}}^{2})  
\end{equation}
where N is the number of lines per element. The uncertainties $\sigma$ for each element are collected in Tab.~\ref{uncer}.\\
\end{appendix}

\begin{acknowledgements} 
RM thanks Pr. Charles Cowley and  Pr. David Gray for their insightful comments during the analysis of HR 8844.
We thank the OHP night assistants for their helpful support during the three observing runs.
This work has made use of the VALD database \citep{VALD}, operated at Uppsala University, the Institute of Astronomy RAS in Moscow, and the University of Vienna.
We have also used the NIST Atomic Spectra Database (version 5.4) available http://physics.nist.gov/asd.
We also acknowledge the use of the ELODIE archive at OHP available at http://atlas.obs-hp.fr/elodie/.
\end{acknowledgements}

\bibliography{ref}
\bibliographystyle{aa}

\newpage
\begin{deluxetable}{lllrll}
\tabletypesize{\normalsize}  
\tablecaption{Line identifications for HR 8844}
\tablecolumns{5}\tablewidth{0pc}  
\tablehead{$\lambda_{\mathrm{obs}}$ (\AA)  & $\lambda_{\mathrm{lab}}$ (\AA) & Identification & $\log gf$ & $E_{low}$ & Comments \\}
\startdata
3900.55     & 3900.515    & \ion{Fe}{1}    &  -0.92   & 26140.176 &  \\
3900.55     & 3900.550    & \ion{Ti}{2}    &  -0.45   & 9118.260  &  \\
3903.08     & 3902.945   & \ion{Fe}{1} & -0.47 & 12560.933 & \\
3903.99     & 3903.99    & \ion{Fe}{1} & -2.53 & 34121.602 & \\
3905.58     & 3905.525  & \ion{Si}{1}     &  -1.09  & 15394.369  & \\
3905.78     & 3905.644  & \ion{Cr}{2}   &  -0.90  & 42986.619  &  \\
3906.18     & 3906.035  & \ion{Fe}{2}   &  -1.83  & 44929.549  &    \\
3920.44     & 3920.26   & \ion{Fe}{1} & -1.75 & 978.074  &    \\
\enddata
\end{deluxetable}

\newpage
\begin{deluxetable}{lcccrcc}
\tabletypesize{\normalsize}   
\tablecaption{Elemental Abundances from unblended lines for HR 8844. References are defined in Tab.~5.}
\tablecolumns{7}\tablewidth{0pc}  
\tablehead{Element  & $\lambda$ (\AA) & $\log gf$ & Ref. & $\log(X/H)_{HR8844}$  & $\log(X)_{\odot}$ &$\log(X)_{\odot} - 12 $  \\}
\startdata
\ion{He}{1} & 4471.470    &-2.211  &   NIST/ASD      	& -1.37 &10.93&-1.07   \\
\ion{He}{1} & 4471.474    &-0.287 &   NIST/ASD     & -1.37 &&  \\
\ion{He}{1} & 4471.474    & -1.035 &  NIST/ASD      & -1.37 && \\
\ion{He}{1} & 4471.485    & -1.035 & NIST/ASD	& -1.37  &&  \\
\ion{He}{1} & 4471.489    & -0.558 & NIST/ASD	& -1.37 &&   \\
\ion{He}{1} & 4471.683    & -0.910  & NIST/ASD	& -1.37 &&\\
\ion{He}{1} & 5875.598    & -1.516  & NIST/ASD      &  -1.22&& \\
\ion{He}{1} & 5875.613    & -0.339  & NIST/ASD	&  -1.22&& \\
\enddata
\label{abundHR8844}
\end{deluxetable}


\newpage
\begin{deluxetable}{lccccr}
\tabletypesize{\normalsize}   
\tablecaption{Elemental Abundances for the comparison stars $\nu$ Cap, $\nu$ Cnc, Sirius A and HD 72660}
\tablecolumns{5}\tablewidth{0pc}  
\tablehead{Element  & $\lambda$ (\AA) & $\log(X/H)_{\nu Cap}$ & $\log(X/H)_{\nu Cnc}$  & $\log(X/H)_{Sirius}$ &  $\log(X/H)_{HD72660}$  \\}
\startdata
\ion{He}{1} & 4471.470    &  -1.07   & -1.47  & -1.77   & -2.07 \\
\ion{He}{1} & 4471.474    & -1.07  & -1.47    &  -1.77 & -2.07  \\
\ion{He}{1} & 4471.474    & -1.07  &  -1.47 & -1.77 & -2.07 \\
\ion{He}{1} & 4471.485    & -1.07  & -1.47 & -1.77  &  -2.07 \\
\ion{He}{1} & 4471.489    &  -1.07  & -1.47 & -1.77   & -2.07   \\
\ion{He}{1} & 4471.683    &  -1.07  &  -1.47 & -1.77 &-2.07  \\
\ion{He}{1} & 5875.598    & -1.07   &  -1.47 & -1.77  & -2.07  \\
\ion{He}{1} & 5875.613    & -1.07  &  -1.47 & -1.77  & -2.07\\
\enddata
\label{other_stars}
\end{deluxetable}
%

\newpage
\begin{table}
\centering
\caption{Abundance uncertainties for the elements analysed in  HR 8844}
\begin{tabular}{cccccccccc}


\hline
  && HeI & CII&OI&MgII&AlII&SiII &SII&CaII \\ \hline
$\Delta$ \teff  & +200 K & -0.30  & 0.00  & 0.05    & 0.047 &0.075  & -0.017  &  0.00 &  0.109\\
$\Delta \log g$ &  0.15 dex& 0.079  & 0.24  & -0.06   & -0.052   & 0.118  & -0.017    & -0.12 & -0/032\\
$\Delta \xi_{t}$ & +0.20 \kms & 0.00 & 0.00  & 0.00   & 0.00 &  0.075  & -0.035   & -0.046 & 0.00 \\
$\Delta \log gf $ & +0.10     & -0.30 & -0.12  & -0.16  & -0.25 & -0.028   & -0.053 &  -0.071 & -0.105  \\
$\Delta continuum$ &          & -0.12 & 0.21  & 0.075    & 0.13 & 0.25  & 0.062    & 0.061 & 0.058  \\
\hline      
$\sigma_{[X/H]}$  & & 0.45  & 0.34 &  0.19 & 0.29  & 0.20 & 0.092 & 0.159 & 0.165 \\ 
  \hline
&  &ScII&TiII&VII&CrII&MnII&FeII&NiII&SrII\\
\end{tabular}
\end{table}

\begin{deluxetable}{l|c}
\tabletypesize{\normalsize}   
\tablecaption{Bibliographical references for the atomic data.}
\tablecolumns{2}\tablewidth{12cm}  
\tablehead{Element  &   Bibliographical source \\}
\startdata
He I & Ma60=\cite{MA60} \\
He I & DR2006=\cite{DR2006}\\
C II & NIST/ASD=\cite{Kramida2017} \\
O I & KZ91=\cite{KZ91}				\\
O I & W96=\cite{W96}				\\
O I & NIST/ASD=\cite{Kramida2017}				\\
Mg II & NIST/ASD=\cite{Kramida2017}			\\
Mg II & NIST/ASD=\cite{Kramida2017}	\\
Al II & NIST/ASD=\cite{Kramida2017} \\
Al I & NIST/ASD=\cite{Kramida2017}	\\
Si II & NIST/ASD=\cite{Kramida2017} \\
Si II & Sh61=\cite{Sh61} \\
S  II & GUES=\cite{GUES} \\
S  II & KP=\cite{KP} \\
S  II & MWRB=\cite{MWRB} \\
Ca II & NIST="http://physics.nist.gov/PhysRefData/ASD/" \\
Ca II & T89=\cite{T89} \\
Sc II & NIST/ASD=\cite{Kramida2017} \\
Ti II & NIST/ASD=\cite{Kramida2017} \\
Ti II & NIST/ASD=\cite{Kramida2017}  \\
Ti II & NIST/ASD=\cite{Kramida2017} \\
Ti II & NIST/ASD=\cite{Kramida2017} \\
V II & K10=\cite{K10} \\
V II & BGF=\cite{BGF} \\
V II & WLDSC=\cite{WLDSC} \\
Cr II & SN14=\cite{SN14} \\
Mn II & KL01=\cite{KL01} \\
Fe II & NIST/ASD=\cite{Kramida2017} \\
Fe II & RU98=\cite{RU98} \\
Ni II & K03=\cite{K03} \\
Sr II & PBL95=\cite{PBL95} \\
Y II & Bie11=\cite{Bie11} \\
Y II & MCS75=\cite{MCS75} \\
Zr II& L06=\cite{L06} \\
Ba II & NBS=\cite{1969AD......1....1M} \\
Ba II & DSVD92= \cite{DSVD92} \\
La II & Zi99= \cite{Zhiguo1999} \\
Ce II & LSCI=\cite{LSCI}\\
Ce II & PQWB=\cite{PQWB}\\
Pr III& ISAN=\cite{ISAN}\\ 
Nd III& RRKB=\cite{RRKB} \\
Sm II & LA06= \cite{LA06} \\
Eu II & LWHS=\cite{LWHS} \\
Eu II & ZLLZ=\cite{ZLLZ} \\
Gd II & DH06= \cite{DH06} \\
Dy II & WLN=\cite{WLN}\\
Dy II & MC=\cite{MC} \\
Gd II & DH06= \cite{DH06} \\
Tb II & LA01= \cite{Law01} \\
Ho II & LA04= \cite{LA04} \\
Er II & LA08= \cite{LA08} \\
Hg II & SR01=\cite{SR01} \\
Hg II & Do03=\cite{Do03}\\ \hline 
\enddata
\label{atomicbib}
\end{deluxetable}


\begin{deluxetable}{|l|c|c|}
\tabletypesize{\scriptsize}   
\tablecaption{Comparison of the derived abundances for HR~8844 derived in this work with the ones of \cite{Royer}.}
\tablecolumns{9}\tablewidth{15cm}  
\tablehead{Element  & $\log(X/H)_{HR8844}$ this work  &  $\log(X/H)_{HR8844}$ \citep{Royer}  }
\startdata
C&	-3.880&	-3.700 \\
O&	-3.270&	-3.300 \\
Mg&	-4.770&	-4.280 \\
Si&	-4.510&	-4.208 \\
Ca&	-5.830&	-5.823 \\
Sc&	-8.990&	-9.223 \\
Ti&	-6.830&	-7.015 \\ 
Cr&	-6.180&	-6.098 \\
Fe&	-4.500&	-4.466 \\
Sr&	-8.330&	-8.480 \\
Y&	-8.460&	-9.500 \\
Zr&	-8.630&	-8.900
\enddata
\label{hr8844_comp}
\end{deluxetable}

\begin{deluxetable}{|l|c|c||c|c||c|c||c|c|}
\tabletypesize{\scriptsize}   
\tablecaption{Comparison of the derived abundances for $\nu$ Cap, $\nu$ Cnc, Sirius A and HD~72660 with previous analyses. A91 refers to Adelman (1991), SD93 to Smith \& Dworetsky (1993), A89 to Adelman (1989), L11 to Landstreet (2011), C16 to Cowley et al. (2016) and GL16 to Golriz \& Landstreet (2016).}
\tablecolumns{9}\tablewidth{15cm}  
\tablehead{Element  & $\log(X/H)_{\nu Cap}$ & A91 and SD93 & $\log(X/H)_{\nu Cnc}$  &A89& $\log(X/H)_{Sirius}$ & L11 and C16& $\log(X/H)_{HD72660}$& GL16  }
\startdata
He &-1.07 &-1.19 & -1.47& -1.57& -1.77&-1.24& & \\
C&-3.48&-3.69 , -3.39&	-3.78&	-3.88&	-4.04	&-4.55	&-4.15&	-4.10 \\
N&-4.02&-4.26&&&-4.02&-4.08&-4.02&	-4.70\\
O&	-3.02&-3.33&	&&-3.63&	-3.60&	-3.57	&-2.50 \\
Na&&&&&	-5.13&	-4.7&&\\
Mg & -4.42&-4.61&-4.53&-4.71&-4.49&-4.70&-4.38&-4.20 \\
Al&-5.53&-6.03&-5.68&-6.38&-5.72&-5.65&-4.90&-4.90 \\
Si & -4.27 &-4.69 &-4.30&-4.59&-4.15&-4.40&-4.13&-4.20 \\
\enddata
\label{literature}
\end{deluxetable}

\allauthors

\listofchanges

\end{document}